\documentclass[preprint,aps,showpacs,showkeys,amsmath,amssymb,nopreprintnumbers,nofootinbib]{revtex4}
\usepackage{graphicx}
\usepackage{dcolumn}
\usepackage{bm}
\usepackage[dvipdfm,bookmarks=true,bookmarksnumbered=true,bookmarkstype=toc]{hyperref}

\def\Rnum#1{\resizebox{0.5em}{\height}{\uppercase\expandafter{\romannumeral #1}}}

\begin{document}


\title{Hyperons in neutron stars}%

\author{Tetsuya Katayama}
\email{6213701@ed.tus.ac.jp}
\affiliation{%
Department of Physics, Faculty of Science and Technology,\\
Tokyo University of Science, Noda 278-8510, Japan 
}%
\author{Koichi Saito}
\email{koichi.saito@rs.tus.ac.jp}
\altaffiliation[Also at ]{J-PARC Branch, KEK Theory Center, Institute of Particle and Nuclear Studies, KEK, Tokai 319-1106, Japan}
\affiliation{%
Department of Physics, Faculty of Science and Technology,\\
Tokyo University of Science, Noda 278-8510, Japan 
}%

\date{\today}

\begin{abstract}
Using the Dirac-Brueckner-Hartree-Fock approach, the properties of neutron-star matter including hyperons are investigated.  In the calculation, we consider both time and space components of the vector self-energies of baryons as well as the scalar ones.  Furthermore, the effect of negative-energy states of baryons is partly taken into account.  We obtain the maximum neutron-star mass of $2.08\,M_{\odot}$, which is consistent with the recently observed, massive neutron stars.  We discuss a universal, repulsive three-body force for hyperons in matter.
\end{abstract}


\maketitle


Neutron stars may be the most dense and exotic state of nuclear matter, and its core serves as a natural laboratory to investigate the nuclear matter whose density reaches several times higher than the normal nuclear-matter density, $n_B^0$ \cite{Glendenning}.  In fact, the recently observed, massive neutron stars, J1614-2230 (the mass of $1.97\pm0.04 \, M_{\odot}$, $M_{\odot}$: the solar mass) \cite{J16142230} and J0348+0432 ($2.01\pm0.04 \, M_{\odot}$) \cite{J03480432}, have provided important information on the equation of state (EoS) for dense nuclear matter.  

To understand these heavy objects, various nuclear models have been examined, in which relativistic mean-field theory (RMFT) is very popular and has been successfully applied to the dense nuclear matter \cite{MFT}. However, in RMFT, nucleon (N)-nucleon short-range correlations in matter cannot be treated.  In contrast, in the Dirac-Brueckner-Hartree-Fock (DBHF) approach, although the calculation is involved, one can consider the effects of the Pauli exclusion principle and short-range correlations. 

Until now, several groups have performed the DBHF calculations not only in the region around $n_B^0$ but also in matter at higher densities 
(see Refs.\cite{DBHF1,DBHF2,DBHF3,Fuchs1,Fuchs2,Fuchs3,Dalen,without_PW,Poschenrieder,Huber,Jong}).  
However, so far there has not been any relativistic attempt to take account of the degrees of freedom of hyperons (Ys) as well as nucleons in dense matter. Because it is quite interesting to see how hyperons contribute to the EoS and to the maximum mass of neutron stars, it seems very urgent to perform the DBHF calculation for matter including hyperons. 

In this Letter, we study such dense neutron-star matter using the DBHF approach.  Here, we particularly pay attention to the following two points: (1) the space component of vector self-energy of baryon (B), $\Sigma^V_B$, is taken into account, because, although it is certainly small at low density, it is expected to be important in dense matter, (2) as in Refs.\cite{Poschenrieder,Huber,Jong}, we partly consider the effect of negative-energy states of baryons in the Bethe-Salpeter (BS) equation to remove the ambiguity in the relationship between the on-shell T-matrix for baryon-baryon  scattering and the baryon self-energies \cite{Fuchs1,Fuchs2,Fuchs3}. 
Furthermore, when hyperons take place in matter, the effective masses of interacting two baryons become very different from each other, and thus we should treat the baryon-mass difference in the BS equation explicitly.  

We now start with the self-energy of baryon in the rest frame of infinite, uniform nuclear matter.  It is given by 
\begin{equation}
    \Sigma_B(k)=\Sigma_B^{S}(k)-\gamma_0\Sigma_B^0(k)+{\boldsymbol\gamma}\cdot\bm{k}\Sigma_B^V(k),  \label{eq_S}
\end{equation}
where $\bm{k} \, (k)$ is the three (four) momentum of baryon.  Here, $\Sigma_B^{S \, (0) \, [V]}$ is the scalar (zero-th component of vector) [space component of vector] part of baryon self-energy.  Using these self-energies, the effective mass, $M_B^{\ast}$, the effective momentum, $\bm{k}_B^{\ast}$, and the effective energy, $E^{\ast}_B$, in matter are defined by
\begin{equation}
    M_B^{\ast}(k) \equiv M_B+\Sigma_B^S(k),  \ \ \ \ \ 
    \bm{k}^{\ast}_B \equiv \bm{k}[1+\Sigma_B^V(k)],  \ \ \ \ \ 
    E^{\ast}_B(k) \equiv \sqrt{\bm{k}_B^{\ast2}+M_B^{\ast2}(k)},    \label{eq_EsB}
\end{equation}
with $M_B$ being the free baryon mass. 
Then, the baryon spinor states with positive or negative energy are respectively constructed as 
\begin{eqnarray}
    \Phi_B(\bm{k},s)&=&\sqrt{M_B^{\ast}(k)+E^{\ast}_B(k)}\left(
    \begin{array}{c}
        \chi_s ,\\
        \frac{\bm{k}_B^{\ast}\cdot{\boldsymbol\sigma}}{M_B^{\ast}(k)+E_B^{\ast}(k)}\chi_{s}
    \end{array}
    \right)\label{eq_positive_spinor},\\
    \Theta_B(\bm{k},s)&=&\sqrt{M_B^{\ast}(k)+E^{\ast}_B(k)}\left(
    \begin{array}{c}
        \frac{\bm{k}_B^{\ast}\cdot{\boldsymbol\sigma}}{M_B^{\ast}(k)+E_B^{\ast}(k)}\chi_{-s} \\
        \chi_{-s}
    \end{array}
    \right)\label{eq_negative_spinor},
\end{eqnarray}
where ${\boldsymbol \sigma}$ is the Pauli matrix, and $\chi_s$ denotes a 2-component Pauli spinor. 

In the conventional DBHF calculation, the baryon-baryon scattering is usually evaluated in the center of mass frame with respect to the interacting two baryons.  In such cases, instead of Eqs.(\ref{eq_positive_spinor})-(\ref{eq_negative_spinor}), the helicity spinors and the partial-wave decomposition are often used to solve the BS equation \cite{DBHF1,DBHF2,DBHF3,Fuchs1,Fuchs2,Fuchs3,Dalen,Poschenrieder,Huber,Jong}.
However, when $\Sigma_B^V$ remains finite and $\bm{k} \neq \bm{k}^{\ast}_B$, although $\bm{k}$ and $\bm{k}^{\ast}_B$ are parallel with each other in the nuclear-matter rest frame, they are not in the center of mass frame.  It is thus more convenient to perform the calculation with the standard spinors, Eqs.(\ref{eq_positive_spinor})-(\ref{eq_negative_spinor}), in the nuclear-matter rest frame, rather than with the helicity spinors in the center of mass frame.

Furthermore, the inclusion of negative-energy states of baryon in the BS amplitude may be necessary to remove the ambiguity of the relationship between the reaction matrices for baryon-baryon scattering and the baryon self-energies \cite{Poschenrieder,Huber,Jong}.  Thus, we here define four reaction amplitudes 
\begin{eqnarray}
    &&T_{B^{\prime\prime\prime}B^{\prime\prime}B^{\prime}B}(\bm{k}^{\prime},\bm{k},s^{\prime\prime\prime},s^{\prime\prime},s^{\prime},s;\bm{P})\nonumber\\
    &\equiv&\bar{\Phi}_{B^{\prime\prime\prime}}\left(\frac{1}{2}\bm{P}+\bm{k}^{\prime},s^{\prime\prime\prime}\right)\bar{\Phi}_{B^{\prime\prime}}\left(\frac{1}{2}\bm{P}-\bm{k}^{\prime},s^{\prime\prime}\right) \, \Gamma \, \Phi_{B^{\prime}}\left(\frac{1}{2}\bm{P}+\bm{k},s^{\prime}\right)\Phi_B\left(\frac{1}{2}\bm{P}-\bm{k},s\right), \label{raT} \\
    &&R_{B^{\prime\prime\prime}B^{\prime\prime}B^{\prime}B}(\bm{k}^{\prime},\bm{k},s^{\prime\prime\prime},s^{\prime\prime},s^{\prime},s;\bm{P})\nonumber\\
    &\equiv&\bar{\Theta}_{B^{\prime\prime\prime}}\left(\frac{1}{2}\bm{P}+\bm{k}^{\prime},s^{\prime\prime\prime}\right)\bar{\Phi}_{B^{\prime\prime}}\left(\frac{1}{2}\bm{P}-\bm{k}^{\prime},s^{\prime\prime}\right) \, \Gamma \, \Phi_{B^{\prime}}\left(\frac{1}{2}\bm{P}+\bm{k},s^{\prime}\right)\Phi_B\left(\frac{1}{2}\bm{P}-\bm{k},s\right),\\
    &&O_{B^{\prime\prime\prime}B^{\prime\prime}B^{\prime}B}(\bm{k}^{\prime},\bm{k},s^{\prime\prime\prime},s^{\prime\prime},s^{\prime},s;\bm{P})\nonumber\\
    &\equiv&\bar{\Phi}_{B^{\prime\prime\prime}}\left(\frac{1}{2}\bm{P}+\bm{k}^{\prime},s^{\prime\prime\prime}\right)\bar{\Phi}_{B^{\prime\prime}}\left(\frac{1}{2}\bm{P}-\bm{k}^{\prime},s^{\prime\prime}\right) \, \Gamma \, \Theta_{B^{\prime}}\left(\frac{1}{2}\bm{P}+\bm{k},s^{\prime}\right)\Phi_B\left(\frac{1}{2}\bm{P}-\bm{k},s\right),\\
    &&P_{B^{\prime\prime\prime}B^{\prime\prime}B^{\prime}B}(\bm{k}^{\prime},\bm{k},s^{\prime\prime\prime},s^{\prime\prime},s^{\prime},s;\bm{P})\nonumber\\
    &\equiv&\bar{\Theta}_{B^{\prime\prime\prime}}\left(\frac{1}{2}\bm{P}+\bm{k}^{\prime},s^{\prime\prime\prime}\right)\bar{\Phi}_{B^{\prime\prime}}\left(\frac{1}{2}\bm{P}-\bm{k}^{\prime},s^{\prime\prime}\right) \, \Gamma \, \Theta_{B^{\prime}}\left(\frac{1}{2}\bm{P}+\bm{k},s^{\prime}\right)\Phi_B\left(\frac{1}{2}\bm{P}-\bm{k},s\right), \label{raP}
\end{eqnarray}
where $\Gamma$ represents the effective reaction operator, and these amplitudes satisfy the following, coupled BS equations
\begin{eqnarray}
    &&T_{BB^{\prime}BB^{\prime}}(\bm{k},\bm{k},s,s^{\prime},s,s^{\prime};\bm{P})=\bar{V}_{BB^{\prime}BB^{\prime}}(\bm{k},\bm{k},s,s^{\prime},s,s^{\prime};\bm{P})\nonumber\\
    &+&\sum_{s^{\prime\prime}s^{\prime\prime\prime}B^{\prime\prime}B^{\prime\prime\prime}}\int\frac{d^3q}{(2\pi)^4}\bar{V}_{BB^{\prime}B^{\prime\prime}B^{\prime\prime\prime}}(\bm{k},\bm{q},s,s^{\prime},s^{\prime\prime},s^{\prime\prime\prime};\bm{P})\nonumber\\
    &&\times Q_{B^{\prime\prime}B^{\prime\prime\prime}}(\bm{P},\bm{q})g_{Th\,B^{\prime\prime}B^{\prime\prime\prime}}(\bm{P},\bm{q})T_{B^{\prime\prime}B^{\prime\prime\prime}BB^{\prime}}(\bm{q},\bm{k},s^{\prime\prime\prime},s^{\prime\prime},s,s^{\prime};\bm{P}),\label{eq_BS_for_T}\\
    &&R_{BB^{\prime}BB^{\prime}}(\bm{k},\bm{k},s,s^{\prime},s,s^{\prime};\bm{P})=\bar{U}_{BB^{\prime}BB^{\prime}}(\bm{k},\bm{k},s,s^{\prime},s,s^{\prime};\bm{P})\nonumber\\
    &+&\sum_{s^{\prime\prime}s^{\prime\prime\prime}B^{\prime\prime}B^{\prime\prime\prime}}\int\frac{d^3q}{(2\pi)^4}\bar{U}_{BB^{\prime}B^{\prime\prime}B^{\prime\prime\prime}}(\bm{k},\bm{q},s,s^{\prime},s^{\prime\prime},s^{\prime\prime\prime};\bm{P})\nonumber\\
    &&\times Q_{B^{\prime\prime}B^{\prime\prime\prime}}(\bm{P},\bm{q})g_{Th\,B^{\prime\prime}B^{\prime\prime\prime}}(\bm{P},\bm{q})T_{B^{\prime\prime}B^{\prime\prime\prime}BB^{\prime}}(\bm{q},\bm{k},s^{\prime\prime\prime},s^{\prime\prime},s,s^{\prime};\bm{P}),\label{eq_BS_for_R}\\
    &&O_{BB^{\prime}BB^{\prime}}(\bm{k},\bm{k},s,s^{\prime},s,s^{\prime};\bm{P})=\bar{W}_{BB^{\prime}BB^{\prime}}(\bm{k},\bm{k},s,s^{\prime},s,s^{\prime};\bm{P})\nonumber\\
    &+&\sum_{s^{\prime\prime}s^{\prime\prime\prime}B^{\prime\prime}B^{\prime\prime\prime}}\int\frac{d^3q}{(2\pi)^4}\bar{V}_{BB^{\prime}B^{\prime\prime}B^{\prime\prime\prime}}(\bm{k},\bm{q},s,s^{\prime},s^{\prime\prime},s^{\prime\prime\prime};\bm{P})\nonumber\\
    &&\times Q_{B^{\prime\prime}B^{\prime\prime\prime}}(\bm{P},\bm{q})g_{Th\,B^{\prime\prime}B^{\prime\prime\prime}}(\bm{P},\bm{q})O_{B^{\prime\prime}B^{\prime\prime\prime}BB^{\prime}}(\bm{q},\bm{k},s^{\prime\prime\prime},s^{\prime\prime},s,s^{\prime};\bm{P}),\label{eq_BS_for_O}\\
    &&P_{BB^{\prime}BB^{\prime}}(\bm{k},\bm{k},s,s^{\prime},s,s^{\prime};\bm{P})=\bar{Z}_{BB^{\prime}BB^{\prime}}(\bm{k},\bm{k},s,s^{\prime},s,s^{\prime};\bm{P})\nonumber\\
    &+&\sum_{s^{\prime\prime}s^{\prime\prime\prime}B^{\prime\prime}B^{\prime\prime\prime}}\int\frac{d^3q}{(2\pi)^4}\bar{U}_{BB^{\prime}B^{\prime\prime}B^{\prime\prime\prime}}(\bm{k},\bm{q},s,s^{\prime},s^{\prime\prime},s^{\prime\prime\prime};\bm{P})\nonumber\\
    &&\times Q_{B^{\prime\prime}B^{\prime\prime\prime}}(\bm{P},\bm{q})g_{Th\,B^{\prime\prime}B^{\prime\prime\prime}}(\bm{P},\bm{q})O_{B^{\prime\prime}B^{\prime\prime\prime}BB^{\prime}}(\bm{q},\bm{k},s^{\prime\prime\prime},s^{\prime\prime},s,s^{\prime};\bm{P}),\label{eq_BS_for_P}
\end{eqnarray}
with $\bar{V}, \bar{U}, \bar{W}$ and $\bar{Z}$ being the anti-symmetrized matrices of one-boson-exchange (OBE) interaction \cite{Gross} with respect to the positive- and negative-energy states (as seen in Eqs.(\ref{raT})-(\ref{raP})).  

In Eqs.(\ref{eq_BS_for_T})-(\ref{eq_BS_for_P}), $Q_{BB^{\prime}}$ is the Pauli exclusion operator for baryons $B$ and $B^\prime$, and $g_{ThBB^{\prime}}$ denotes the  Thompson's two-particle propagator \cite{Thompson_propagator}. 
The seven arguments in the four reaction amplitudes, $T,\,R,\,O,\,P$, are as follows: from left to right, the first variable represents the final (or intermediate) relative three-momentum; the second, the initial (or intermediate) relative three-momentum; the third and fourth are for the spins of the final (or intermediate) two baryons, each of which is up ($+$) or down ($-$); the fifth and sixth, the spins of the initial (or intermediate) two baryons; and the last one is the total three-momentum of interacting two baryons.  We note that the negative-energy states appear only in the initial and/or final states of the BS amplitudes, and they are not included in the intermediate states, because, in the realistic baryon-baryon potentials such as the Bonn potentials, the negative-energy states are usually not considered \cite{Poschenrieder,Huber}. 

The ladder-approximated, coupled BS equations can be numerically solved in the nuclear-matter rest frame.  To reduce the number of variables and make the present calculation feasible, we here average the azimuthal angle in the spinors, Eqs.(\ref{eq_positive_spinor})-(\ref{eq_negative_spinor}), namely we replace $E^{\ast}_B(1/2\bm{P}\pm\bm{k})$ by the averaged one, $\frac{1}{2\pi}\int d\phi E^{\ast}_B(1/2\bm{P}\pm\bm{k})$. We have checked that this change does not lead any large numerical error in our final results. 

Given the reaction amplitudes, we can calculate the following components \cite{Huber}
\begin{eqnarray}
    \Sigma_{\Phi\Phi}^B(k)&\equiv&\bar{\Phi}_B(k,+) \, \Sigma_B(k) \, \Phi_B(k,+)\nonumber\\
    &=& 2M_B^{\ast}(k)\Sigma_B^S(k)-2E^{\ast}_B(k)\Sigma_B^0(k)+2\bm{k}\cdot\bm{k}_B^{\ast}\Sigma_B^V(k) ,\label{eq_SPP}\\
    \Sigma_{\Theta\Phi}^B(k)&\equiv&\bar{\Theta}_B(k,+) \, \Sigma_B(k) \, \Phi_B(k,-)\nonumber\\
    &=& 2|\bm{k}^{\ast}_B|(k)\Sigma_B^0(k)-2|\bm{k}|E_B^{\ast}(k)\Sigma_B^V(k) ,\label{eq_STP}\\
    \Sigma_{\Theta\Theta}^B(k)&\equiv&\bar{\Theta}_B(k,+) \, \Sigma_B(k) \, \Theta_B(k,+)\nonumber\\
    &=& -2M_B^{\ast}(k)\Sigma_B^S(k)-2E^{\ast}_B(k)\Sigma_B^0(k)+2\bm{k}\cdot\bm{k}_B^{\ast}\Sigma_B^V(k),\label{eq_STT}
\end{eqnarray}
where $\Sigma_B(k)$ is given by Eq.(\ref{eq_S}), and the components, $\Sigma_{\Phi\Phi}^B(k), \Sigma_{\Theta\Phi}^B(k), \Sigma_{\Theta\Theta}^B(k)$, are respectively calculated through the reaction amplitudes, $T, R, P$.  Using these relations, we can uniquely determine the self-energies in Eq.(\ref{eq_S}), and calculate the energy density and pressure of matter \cite{DBHF3}.  We here discard the contribution of retardation effect.  

Now we are in a position to show our results. Through the whole calculation, we adopt the Bonn potentials \cite{DBHF1}, and use the conventional ``reference spectrum'' approximation \cite{DBHF1,DBHF2,DBHF3,Fuchs1,Fuchs2,Fuchs3,Dalen,without_PW,Jong}, where the momentum dependence of the self-energies is frozen at some reference momentum.  Here, the reference point is chosen to be the Fermi momentum, $k_{F}$, at each nuclear density, $n_B$. 

\begin{table}
\caption{\label{tab:matter_property}
Calculated properties of symmetric nuclear matter at the saturation point, $n_B^0$.  In the first column, the results of Bonn A, B and C are respectively labeled by A, B and C, while A$^\ast$ denotes the result of Bonn A with the modified coupling $g^\ast_{NN\sigma}$ (see Eq.(\ref{eq:gBBs})).  The values of the binding energy per particle, ${\cal E}/n_B^0-M_N$, the incompressibility, $K$, the symmetry energy, $S$,  and the slope parameter, $L$, are in MeV, and $n_B^0$, is in fm$^{-3}$. The fitted values in A$^\ast$ are denoted by $\dagger$.}
\begin{ruledtabular}
\begin{tabular}{ccccccc}
case&$n_B^0$&${\cal E}/n_B^0-M_N$&$K$&$S$&$L$\\
\hline
A&0.149&$-10.5$&204&28.8&78.6\\
B&0.130&$-7.3$&133&22.7&58.2\\
C&0.112&$-5.2$&87&18.0&42.2\\
A$^{\ast}$&0.168$^{\dagger}$&$-15.3^{\dagger}$&233$^{\dagger}$&33.6$^{\dagger}$&95.0\\
\end{tabular}
\end{ruledtabular}
\end{table}
We first study the symmetric nuclear matter around $n_B^0$, where it consists of only nucleons  interacting through the exchanges of $\sigma,\,\delta,\,\omega,\,\rho,\,\eta$ and $\pi$ mesons. 
In Table~\ref{tab:matter_property}, we present the properties of matter at $n_B^0$.  In the present calculation, the binding per particle in A$\sim$C is relatively shallower than the empirical value.  This tendency is close to the result by Poschenrieder and Weigel \cite{Poschenrieder} because our method resembles their approach. 

We try to adjust the matter properties by assuming that the nucleon-$\sigma$ coupling constant varies as a function of the scalar self-energy $\Sigma^S_N$.  Because baryon is a composite object, the meson-baryon coupling strength may generally depend on the scalar density in matter \cite{Jong,birse,saito,ApJ}.  To take account of such an effect, we suppose that the coupling is expressed as 
\begin{equation}
    g_{NN\sigma}^{\ast} = g_{NN\sigma} \left[1 + \sum_{i=1}^4 \alpha_i \left( \frac{\Sigma^S_N}{M_N} \right)^{\!\! i} \, \right] , \label{eq:gBBs}
\end{equation}
where $g_{NN\sigma}$ is the value in vacuum.  The four parameters, $\alpha_{i=1\sim4}$, are determined so as to reproduce the empirical binding value, $n_B^0$, $K$ and $S$ (see Table~\ref{tab:matter_property}), and we then find $\alpha_1 = -0.36$, $\alpha_2 = -1.69$, $\alpha_3 = -3.07$ and $\alpha_4 = -1.86$ for Bonn A potential.  This modification enhances $g_{NN\sigma}$ by only 2\% at $n_B^0$. The properties of matter in this scheme (A$^\ast$) is also shown in the table.  

We notice that, in the present calculation, the space component of vector self-energy, $\Sigma^V_N$, is certainly small around $n_B^0$, but it grows rapidly at high density and reaches about 0.7 at $n_B = 1.0$ fm$^{-3}$ in symmetric nuclear matter (see Eqs.(\ref{eq_S}) and (\ref{eq_EsB})).  Therefore, it is necessary to take account of the space part explicitly in dense matter.

Next we challenge the calculation of neutron-star matter including hyperons.  In the following calculations, we adopt the scheme A$^\ast$. 

We now have to determine the coupling constants for hyperons.  Using the experimental data of nucleon-hyperon scattering, the hyperon-meson coupling constants have been studied by several groups \cite{Holzenkamp}.  However, due to poor experimental accuracy, the coupling constants cannot be determined without large ambiguities.  Thus, in the present calculation, we determine them (except for the hyperon-$\sigma$ couplings) by using SU(6) symmetry \cite{ApJ,SU6}.  For the hyperon-$\sigma$ interactions, we can use the recent analyses of hypernuclei and hyperon production reactions, which  have suggested that the $\Lambda,\,\Sigma^{-}$ and $\Xi^{-}$ may respectively feel the optical potential, $U_{\Lambda^{-},\Sigma^{-},\Xi^{-}}\simeq-27,\,+30,\,-15$ MeV, in nuclear matter \cite{hyppot}.  We thus choose the coupling constants, $g_{YY\sigma}$, so as to reproduce these potential depths at $n_B^0$, using the Schr\"{o}edinger-equivalent optical potential 
\begin{equation}
    U_Y(k) = \Sigma^S_Y(k)-\frac{\Sigma^0_Y(k)}{M_Y} [E^{\ast}_Y(k)-\Sigma^0_Y(k)] + \frac{1}{2M_Y} [ (\Sigma_Y^{S}(k))^2 - (\Sigma^{0}_Y(k))^2 ].
\end{equation}

Furthermore, for nonstrange mesons, a cutoff parameter in the form factor at hyperon-meson vertex, $\Lambda_{YY^{\prime}M}$, is assumed to be the same as in the nucleon-meson form factor, while, for strange mesons, a cutoff parameter, $\Lambda_{BB^{\prime}K}$ ($\Lambda_{BB^{\prime}K^{\ast}}$), is taken to be  the average value of $\Lambda_{BB^{\prime}\eta}$ and $\Lambda_{BB^{\prime}\pi}$ ($\Lambda_{BB^{\prime}\omega}$ and $\Lambda_{BB^{\prime}\rho}$).  

\begin{figure}[htbp]
 \begin{center}
    \includegraphics[width=350pt,keepaspectratio,clip,angle=270]{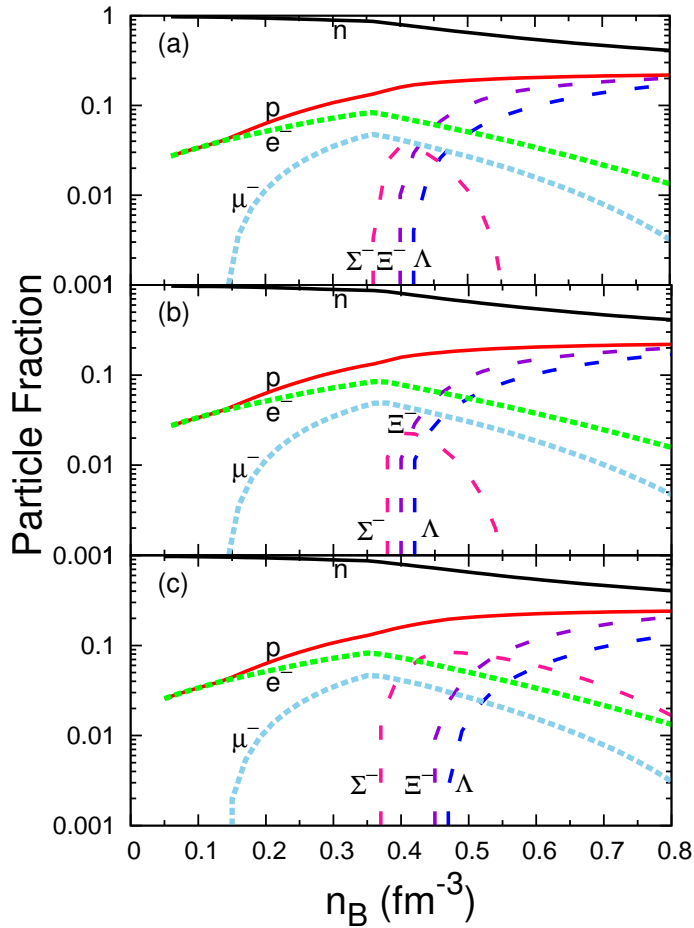}
 \end{center}
 \caption{(Color online) Particle fractions for (a) B5M6, (b) B8M6 and (c) B5M8.}
 \label{fig_B5_B8_B5K}
\end{figure}
Because an enormous amount of time is needed to perform the full calculation, we first perform two preliminary calculations: one includes $e^-$, $\mu^-$ and five baryons (neutron (n), proton (p), $\Lambda$, $\Sigma^-$, $\Xi^-$),\footnote{From among the members of the $\Sigma$ and $\Xi$ hyperons, we select the $\Sigma^-$ and $\Xi^-$ only.  The reason is because, from the viewpoint of electric charge, it is expected that they can appear easier in matter rather than the other members ($\Sigma^+$, $\Sigma^0$ and $\Xi^0$) \cite{ApJ}.} and the other includes the leptons and eight baryons (n, p, $\Lambda$, $\Sigma^-$, $\Sigma^0$, $\Sigma^+$, $\Xi^-$, $\Xi^0$).  In these calculations, we consider only six nonstrange mesons ($\sigma,\,\delta,\,\omega,\,\rho,\,\eta,\,\pi$), and exclude the baryon-exchange and baryon-transition processes such as $N + \Lambda \to \Lambda + N$, $N + \Lambda \to N + \Sigma$, etc.  Note that these processes are induced by the exchanges of $K$, $K^\ast$ and iso-vector, nonstrange mesons.  We call the first set B5M6, and the second B8M6. 

In the panels (a) and (b) of Fig.\ref{fig_B5_B8_B5K}, we present the particle fractions in B5M6 and B8M6. 
As seen in the figure, both the results are very similar to each other, and show that the $\Sigma^-$ first appears around $n_B\simeq0.37\,\mathrm{fm}^{-3}$, and next the $\Xi^-$ and $\Lambda$ are created.  However, the fraction of $\Sigma^-$ dwindles rapidly with increasing $n_B$.  In contrast, the numbers of $\Xi^-$ and $\Lambda$ grow steadily, once they emerge in matter. 

Because the difference between B5M6 and B8M6 is expected to be small, we proceed to the final calculation, where the five baryons are considered and they interact through the exchanges of eight mesons ($\sigma,\,\delta,\,\omega,\,\rho,\,\eta,\,\pi,\,K,\,K^{\ast}$).  We here include the effect of the baryon-exchange and baryon-transition processes.  We call this scheme B5M8.  
In the panel (c) of Fig.\ref{fig_B5_B8_B5K}, the particle fraction for B5M8 is displayed.  It is interesting to notice that the result is again similar to the previous ones, but the fraction of $\Sigma^-$ is enhanced by the baryon-transition process between $\Sigma^-$ and $\Lambda$ hyperons.  Furthermore, comparing with the result in B5M6, the thresholds for the $\Lambda$ and $\Xi^{-}$ in B5M8 move toward higher density. 

\begin{figure}[htbp]
 \begin{center}
    \includegraphics[width=310pt,keepaspectratio,clip,angle=270]{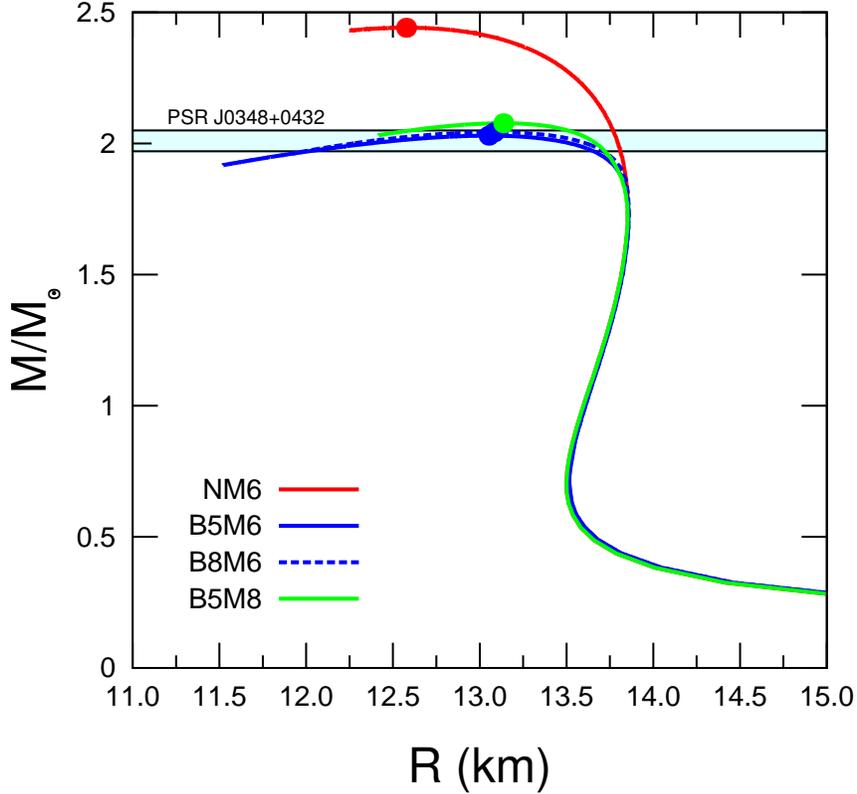}
 \end{center}
 \caption{(Color online) Mass-radius relations for neutron stars.
The dot on each line represents the maximum mass (see also Table \ref{tab:NS_NlY_property}). The shaded area represents the mass of J0348+0432.}
 \label{fig_NS_NlY}
\end{figure}
\begin{table}
\caption{\label{tab:NS_NlY_property}
Neutron-star radius, $R_{max}$ (in km), the central density, $n_c$ (in fm$^{-3}$), and the ratio of the maximum neutron-star mass to the solar mass, $M_{max}/M_{\odot}$.}
\begin{ruledtabular}
\begin{tabular}{cccc}
case&$R_{max}$&$n_c$&$M_{max}/M_{\odot}$\\
\hline
NM6&12.6&0.78&2.44\\
B5M6&13.1&0.76&2.03\\
B8M6&13.1&0.75&2.04\\
B5M8&13.1&0.74&2.08\\\hline
\end{tabular}
\end{ruledtabular}
\end{table}
Using the Tolman-Oppenheimer-Volkoff (TOV) equation \cite{TOV} with the BPS model \cite{BPS} for the EoS in the crust region, we can calculate the neutron-star mass as a function of its radius.  The calculation is performed under the conditions of charge neutrality and $\beta$-equilibrium in weak interaction.  The present results are summarized in Fig.\ref{fig_NS_NlY} and Table \ref{tab:NS_NlY_property}, where we show the mass-radius relations and the properties of neutron stars at the maximum mass.  Here, NM6 denotes the result in which only the leptons, nucleons and six nonstrange mesons participate.  We can find that the predicted maximum mass in each case is consistent with the observed ones, $1.97\pm0.04 \, M_{\odot}$ (J1614-2230) \cite{J16142230} and $2.01\pm0.04 \, M_{\odot}$ (J0348+0432) \cite{J03480432}.  

We would like to compare the present results with those in the non-relativistic Brueckner-Hartree-Fock (BHF) approach.  It is well recognized that the saturation properties of symmetric nuclear matter cannot be explained by the BHF calculation with two-body interactions, and that it is vital to consider a repulsive three-body force (TBF) additionally to move the calculated saturation point toward the empirical value \cite{kohno}.  In contrast, the DBHF calculation can provide a result close to the empirical value without further ingredients.  This is because the DBHF approach involves an inherent ability to account for important TBFs through its density dependence, i.e., the TBF originating from virtual excitation of a nucleon-antinucleon pair, known as $Z$-graphs \cite{DBHF1}.  In fact, it is not difficult to find the term of $Z$-graphs, $\Delta E_{pair}^N \approx {\vec k}^{\, 2} (\Sigma_N^S)^2 / 2M_N^3$, in the nucleon energy at the mean-field level \cite{cohen,wallace,saito2}.  The relativistic effect on the binding energy per nucleon is then well fitted as $\Delta ({\cal E}/n_B^0)_{rel} \propto (n_B / n_B^0)^{8/3}$ \cite{DBHF1}, which depends on $n_B$ strongly, and it helps obtain a better saturation point \cite{samm}. From this point of view, the contribution of $Z$-graphs plays an important role in success of Dirac phenomenology.  On the other hand, some people have argued that the pair creation should be suppressed by the compositeness of nucleon \cite{cohen,seki}.\footnote{In Ref.\cite{cohen}, it has been emphasized that the ``$Z$-graphs'' in Dirac phenomenology should {\it not} be interpreted as arising from virtual nucleon-antinucleon pairs.}  However, even when the effect of the compositeness is taken into account, the repulsive term still remains \cite{saito2,birse}. 

\begin{figure}[htbp]
 \begin{center}
    \includegraphics[width=300pt,keepaspectratio,clip,angle=270]{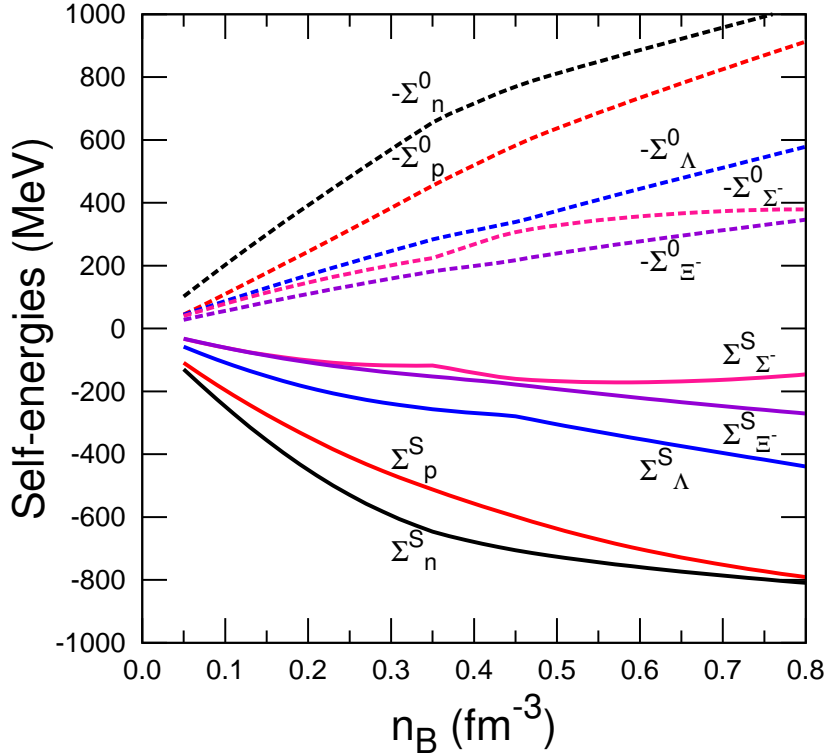}
 \end{center}
 \caption{(Color online) Self-energies, $\Sigma_B^S$ and $\Sigma_B^0$, in B5M8.}
 \label{fig_self_energies}
\end{figure}

In dense neutron-star matter, where hyperons can also participate, the similar situation may occur.  In this case, it is again well known that, in the BHF approach, the inclusion of hyperons softens the EoS very much, and that such an EoS is  inconsistent with the existence of heavy neutron stars \cite{schulze}.  To remedy this problem, it may be again necessary to introduce repulsive TBFs for hyperons \cite{NBHF}.  On the other hand, the DBHF calculation inherently contains the density-dependent, repulsive TBF, and it seems to be {\it universal} for all baryons \cite{saito2,birse}.  As long as the magnitudes of $\Sigma_B^S$ and $\Sigma_B^0$ are large, due to Lorentz structure, each baryon feels the  repulsive potential, $\Delta E_{pair}^B$, in nuclear matter, which may again play an important part in obtaining the EoS for sustaining the massive neutron stars, as shown in the present calculation.  In Fig.\ref{fig_self_energies}, we show the self-energies  for hyperons as well as nucleons.  We can see that their magnitudes are certainly of the order of 100MeV at high densities. 

In summary, using the Dirac-Brueckner-Hartree-Fock approach, we have studied the properties of neutron-star matter including hyperons.  The result has shown that the $\Sigma^-$, $\Lambda$ and $\Xi^-$ appear in dense matter, but the fraction of $\Sigma^-$ decreases with increasing $n_B$. The maximum neutron-star mass is estimated to be $2.08\,M_{\odot}$, which is consistent with the masses of heavy neutron stars.  Thus, we can conclude that it is very important to consider not only the effects of Pauli exclusion principle and short-range correlations in matter but also the relativistic effect involved in Dirac phenomenology.  In the present calculation, we have determined the hyperon-meson couplings by SU(6) symmetry.  However, those couplings should be improved in the future calculation. 
In the DBHF approach, it is very difficult to understand a neutron star with heavier mass than that of J0348+0432.  Thus, if such an object is found in the future, it may clearly suggest that the nonbaryonic degrees of freedom (like quarks) emerge in its core region.

\vspace{1cm}
\begin{acknowledgements}
This work was supported by JSPS KAKENHI Grant Number 255742.
\end{acknowledgements}

%
%

%

\end{document}